\documentclass[11pt]{article}
\input{psfig.sty}
\usepackage{amsmath}
\usepackage{amssymb}
\usepackage{times}
\begin{document}
\pagestyle{plain}
\hsize = 6.5 in 				
\vsize = 8.5 in		   
\hoffset = -0.75 in
\voffset = -0.5 in
\baselineskip = 0.31 in

\begin{center}
Accepted for publication in

Biophysical Chemistry Special Issue for Walter Kauzmann

\vskip 1.cm 
{\Large\bf Thermodynamic and Kinetic Analysis of Sensitivity}

\vskip 0.15cm
{\Large\bf Amplification in Biological Signal Transduction}
\vskip 0.5 true cm 
Hong Qian
\vskip 0.5 true cm 
Department of Applied Mathematics, University of
Washington\\
Seattle, WA 98195-2420
\vskip 0.3cm
\today
\vskip 1.0 true cm
\end{center}

\begin{quote}
{\bf Based on a thermodynamic analysis of the kinetic model
for the protein phosphorylation-dephosphorylation cycle, we 
study the ATP (or GTP) energy utilization of this ubiquitous 
biological signal transduction process.
It is shown that the free energy from hydrolysis
inside cells, $\Delta G$ (phosphorylation potential),
controls the amplification and sensitivity of the 
switch-like cellular module; the response coefficient of the 
sensitivity amplification approaches the optimal 1 and the
Hill coefficient increases with increasing $\Delta G$.  We 
discover that zero-order ultrasensitivity is mathematically 
equivalent to allosteric cooperativity.  Furthermore, we 
show that the high amplification in ultrasensitivity 
is mechanistically related to the proofreading kinetics 
for protein biosynthesis.  Both utilize multiple kinetic 
cycles in time to gain temporal cooperativity, in contrast to 
allosteric cooperativity that utilizes multiple subunits 
in a protein. 
}
\end{quote}

\pagebreak

\section{Introduction}

Biological signal transduction processes are increasingly being 
understood in quantitative and modular terms \cite{K,HHLM}.  One 
of the most commonly studied modules of cellular ``circuitry'' 
is the phosphorylation-dephosphorylation cycle (PdPC) \cite{Krebs} 
which has been shown to exhibit sensitivity amplification for the 
appropriate stimuli expressed through activating a kinase or 
inhibiting a phosphatase \cite{SC,GK,KGS}.  Both experimental 
measurement \cite{SCS,HF,FM} and theoretical modeling have shown 
that the covalent modification gives rise to a switch-like 
behavior.

	Sensitivity amplification requires energy consumption 
\cite{SC,SCS,GK2}.  Since the PdPC involves the transfer of 
high-energy phosphate group, it is natural to ask how the 
cellular phosphoenergetics play a role in the signal 
transduction processes.  Recently, a novel mechanism has 
been proposed \cite{LQ} for improved Rab 5 GTPase function 
as cellular timer \cite{BSM} by utilizing the energy derived 
from GTP hydrolysis.  It is shown that an energy expenditure 
is necessary for a GTPase timer to be accurate and robust. 

	Phosphoenergetics and ATP hydrolysis are also involved in
PdPC.  While it is known that energy expenditure is required
to maintain levels of phosphorylation in excess of an equilibrium
\cite{SC,GK2}, it is still not yet clear how cellular energetics 
relates to this type of signal transduction process.  One approach 
to address this question is introducing a rigorous thermodynamic 
analysis into the kinetic models of PdPC \cite{GK,HF}.  The 
simplest kinetic scheme for PdPC is shown in (\ref{rxn}), which is 
based on a model proposed by Stadtman and Chock \cite{SC} and by
Goldbeter and Koshland \cite{GK}.  The essential difference between 
our (\ref{rxn}) and the earlier models is the nonzero $q_1$ 
and $q_2$, i.e., the reversibility of the separate and 
distinct phosphorylation and dephosphorylation processes. 

	In order to carry out a cogent thermodynamic analysis for
the kinetic model of PdPC, the reversibility of the biochemical 
reactions involved, specifically the phosphorylation catalyzed by 
kinase and dephosphorylation catalyzed by phosphatase, must be 
enforced.  While this was known to be an important issue
\cite{Gresser}, almost all current models neglect the slow reverse 
steps.

\section{Basic Biochemical Equilibrium and Energetics}

	We consider a phosphorylation-dephosphorylation cycle 
(PdPC) catalyzed by kinase $E_1$ and phosphatase $E_2$ respectively.  
The phosphorylation covalently modifies the protein $W$ 
to become $W^*$:
\begin{eqnarray}
   W + E_1 
   \overset{a_1}{\underset{d_1}{\rightleftharpoons}} 
   &WE_1& 
   \overset{k_1}{\underset{q_1}{\rightleftharpoons}} 
   W^* + E_1  \hspace{1in} (I)
\nonumber\\
\label{rxn}\\[-12pt]
   W^* + E_2 
   \overset{a_2}{\underset{d_2}{\rightleftharpoons}} 
   &W^*E_2& 
   \overset{k_2}{\underset{q_2}{\rightleftharpoons}} 
   W + E_2.  \hspace{1in}  (II)
\nonumber
\end{eqnarray}
It is important to note that the reaction I
is not the reverse reaction of II.  In fact, recognizing
that the hydrolysis reaction ATP $\rightleftharpoons$ ADP+Pi
explicitly, we have 
\begin{eqnarray*}
   W + E_1 + ATP
   \overset{a_1^o}{\underset{d_1}{\rightleftharpoons}} 
   &W\cdot E_1\cdot ATP& 
   \overset{k_1}{\underset{q_1^o}{\rightleftharpoons}} 
   W^* + E_1 + ADP 
\\
   W^* + E_2 
   \overset{a_2}{\underset{d_2}{\rightleftharpoons}} 
   &W^*E_2& 
   \overset{k_2}{\underset{q_2^o}{\rightleftharpoons}} 
   W + E_2 + Pi.
\end{eqnarray*}
Thus, at constant concentrations for ATP, ADP, and Pi,
\begin{equation}
            a_1 = a_1^o[ATP], \hspace{0.5cm}
            q_1 = q_1^o[ADP], \hspace{0.5cm}
	    q_2 = q_2^o[Pi].
\end{equation}
For simplicity we have assumed that these rate constants
are pseudo-first order, which implies that ATP, ADP,
and Pi are sufficiently below the saturation levels for 
their respective enzymes.  

The equilibrium constant for ATP hydrolysis therefore is 
\begin{equation}
        \frac{[ATP]_{eq}}{[ADP]_{eq}[Pi]_{eq}} =
	        \frac{d_1q_1^od_2q_2^o}{a_1^ok_1a_2k_2}
               = e^{-\Delta G^o/RT},
\end{equation}  
where $\Delta G^o$ is the standard free-energy change for
ATP hydrolysis reaction \cite{Str}.  That is 
$\frac{a_1k_1a_2k_2}{d_1q_1d_2q_2}$ = 1 in
equilibrium. However, with physiological concentrations for 
ATP, ADP, and Pi inside cells, the quotient 
\begin{equation}
        \gamma = \frac{a_1k_1a_2k_2}{d_1q_1d_2q_2} 
	       = \frac{a_1^ok_1a_2k_2}{d_1q_1^od_2q_2^o} 
		\left(\frac{[ATP]}{[ADP][Pi]}\right),
\label{eq4}
\end{equation}
is directly related to the intracellular phosphorylation
potential
\begin{equation}
              RT\ln\gamma  
		= \Delta G^o+ RT\ln\frac{[ATP]}{[ADP][Pi]}
                = \Delta G 
\end{equation}
where $RT=0.6kcal/mol$ at room temperature.
We shall also introduce an equilibrium constant for the
dephosphorylation reaction catalyzed by phosphatase under
intracellular phosphate concentration:
\begin{equation}
                  \mu = \frac{d_2q_2}{k_2a_2}.
\end{equation}
The two parameters $\gamma$ and $\mu$ are the key augmentations
to the model of Goldbeter and Koshland \cite{GK}.

	We recognize the fact that there is currently no 
experimental evidence for reaction II being reversible.  While 
the  backward rate for the dephosphorylation reaction catalyzed 
by a phosphatase can be extremely small, a thermodynamically
correct model has to have a nonzero $q_2^o$, no matter how small 
it is.  In fact, Eq. \ref{eq4} could be used to estimate
the unmeasurable $q_2^o$ if all the other rate constants 
are known.

\section{Reversible Kinetic Model for Covalent Modification}

	The kinetic equations for the reaction cycle in
(\ref{rxn}) are straightforward
\begin{eqnarray}
  \frac{d[W]}{dt} &=& -a_1[W][E_1]+d_1[WE_1]+k_2[W^*E_2]-q_2[W][E_2]
\nonumber\\[10pt]
  \frac{d[WE_1]}{dt} &=& a_1[W][E_1]-(d_1+k_1)[WE_1]+q_1[W^*][E_1]
\nonumber\\
\label{mastereq}\\[-5pt]
 \frac{d[W^*]}{dt} &=& -a_2[W^*][E_2]+d_2[W^*E_2]+k_1[WE_1]-q_1[W^*][E_1]
\nonumber\\[10pt]
 \frac{d[W^*E_2]}{dt} &=& a_2[W^*][E_2]-(d_2+k_2)[W^*E_2]+q_2[W][E_2].
\nonumber
\end{eqnarray}
These equations are solved in conjunction with conservation
equations \ref{wt}, \ref{e1t}, and \ref{e2t}:
\begin{eqnarray}
         W_T &=& [W]+[W^*]+[WE_1]+[W^*E_2]
\label{wt}\\
         E_{1T} &=& [E_1]+[WE_1]
\label{e1t}\\
          E_{2T} &=& [E_2]+[W^*E_2].
\label{e2t}
\end{eqnarray}
Following the elegant mathematical treatment given in \cite{GK}, 
we have the steady-state fraction of phosphorylated $W$,
denoted by $W^*=[W^*]/W_T$ as in \cite{GK}, satisfying 
\begin{equation}
          \sigma = \frac{\mu\gamma\left[\mu-(\mu+1)W^*\right]
		(W^*-K_1-1)}
	{\left[\mu\gamma-(\mu\gamma+1)W^*\right](W^*+K_2)}. 
\label{sigma}
\end{equation}
Here we have denoted
\[      \sigma = \frac{k_1E_{1T}}{k_2E_{2T}}, \hspace{0.4cm}
	K_1 = \frac{d_1+k_1}{a_1W_T}, \hspace{0.4cm}
        K_2 = \frac{d_2+k_2}{a_2W_T}. 
\]
These three parameters are in the original model \cite{GK}
which, by assuming {\it irreversible} reactions with $q_1=q_2=0$,
has $\mu=0$ and $\gamma=\infty$.  $\sigma$ represents the
ratio of kinase activity to phosphatase activity. Hence
it characterizes the magnitude of the stimuli for the PdPC.
$1/K_1$ and $1/K_2$ are the ratios of substrate 
concentrations to the Michaelis-Menten constants 
of kinase and phosphatase, respectively.  A small $K$ ($\ll 1$)
means the enzymatic reaction is highly saturated. 

	More explicitly, Eq.  \ref{sigma} is a quadratic equation
for $W^*$:
\begin{equation}
           AW^{*2} - BW^* + C = 0,
\label{quadratic}
\end{equation}
in which
\begin{eqnarray*}
             A &=& 1+\mu-\sigma\left(1+\frac{1}{\gamma\mu}\right)
\\
             B &=& \mu+(1+\mu)(1+K_1)- 
	\sigma\left(1-K_2\left(1+\frac{1}{\gamma\mu}\right)\right)
\\
             C &=& \mu(1+K_1)+\sigma K_2.
\end{eqnarray*}
The steady-state solution to Eq. \ref{mastereq}, therefore, is the
positive root of Eq. \ref{quadratic}
\begin{equation}
    W^* = \frac{B-\sqrt{B^2-4AC}}{2A}.
\end{equation}
It is plotted in Fig. 1 using $K_1=K_2=0.01$, i.e., both
enzymes are highly saturated thus the rates are only weakly
dependent on the respective substrate concentrations \cite{GK}, 
and $\mu=10^{-3}$, i.e., the dephosphorylation reaction is highly
irreversible \cite{Gresser}.  It is seen that the quality 
of the amplifier is directly related to the phosphorylation 
potential.  In fact, when $\gamma=1$, i.e., ATP 
$\rightleftharpoons$ ADP+Pi are in chemical equilibrium, 
\begin{equation}
       W^* = \frac{a_1k_1}{a_1k_1+d_1q_1}
           = \frac{d_2q_2}{a_2k_2+d_2q_2}
           = \frac{\mu}{1+\mu}
\end{equation}
which is independent of $\sigma$.  In this case, the
amplification is completely abolished.  Biological amplification 
needs energy, just like a home stereo.	
\begin{figure}[b]
\[
\psfig{figure=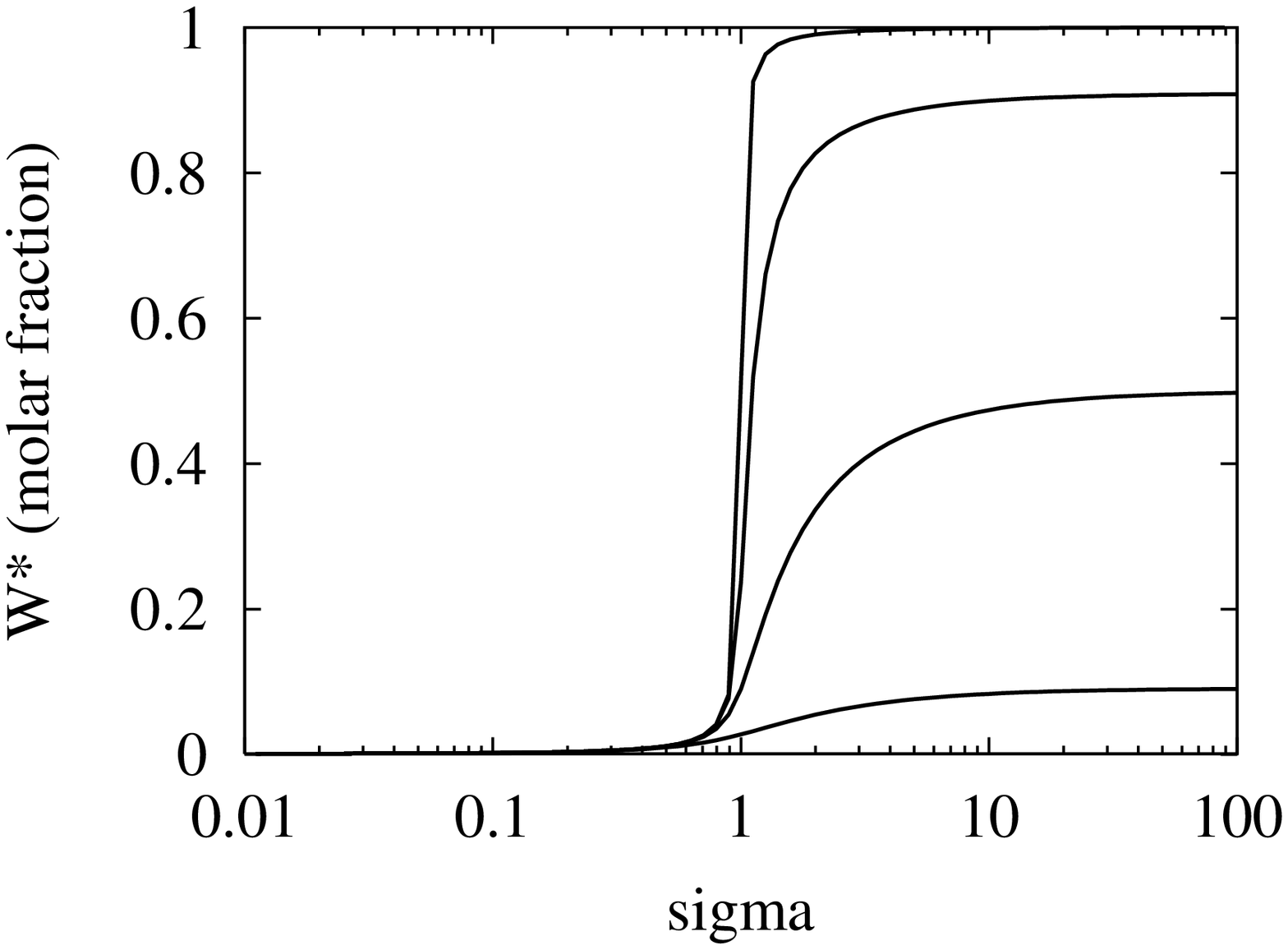,%
width=4.in,height=3.0in,%
bbllx=2.in,bblly=1.in,%
bburx=8.in,bbury=8in,%
angle=-90}
\]
\caption{Amplified sensitivity of a phosphorylation-dephosphorylation
cycle as a function of intracellular phosphorylation potential
$\Delta G$.  From top to bottom, $\gamma$ = $10^{10}$, $10^4$,
$10^3$ and $10^2$, corresponding to $\Delta G$= 13.8, 5.5, 4.1, 
and 2.8 $kcal/mol$. $13.8 kcal/mol$ is the typical value for
intracellular phosphorylation potential \cite{Str,Kush}. Other 
parameters used in the computation: $K_1=K_2=0.01$ and $\mu=0.001$.}
\end{figure}

	The switch-like behavior in Fig. 1 can be understood 
semi-quantitatively as follows (Fig. 2).  The kinase catalyzed 
phosphorylation reaction has a Michaelis-Menten constant
$K_1W_T$ and $V_{max}=V_1=k_1E_{1T}$.  Therefore the overall rate 
of the reaction is $\frac{V_1}{1+K_1}$; similarly 
the dephosphorylation reaction has a rate $\frac{V_2}{1+K_2}$
where $V_2=k_2E_{2T}$.  The equilibrium constants for 
the respective reactions are $\mu\gamma=\frac{a_1k_1}{d_1q_1}$
and $\mu=\frac{d_2q_2}{a_2k_2}$.
\begin{figure}[t]
\[
\psfig{figure=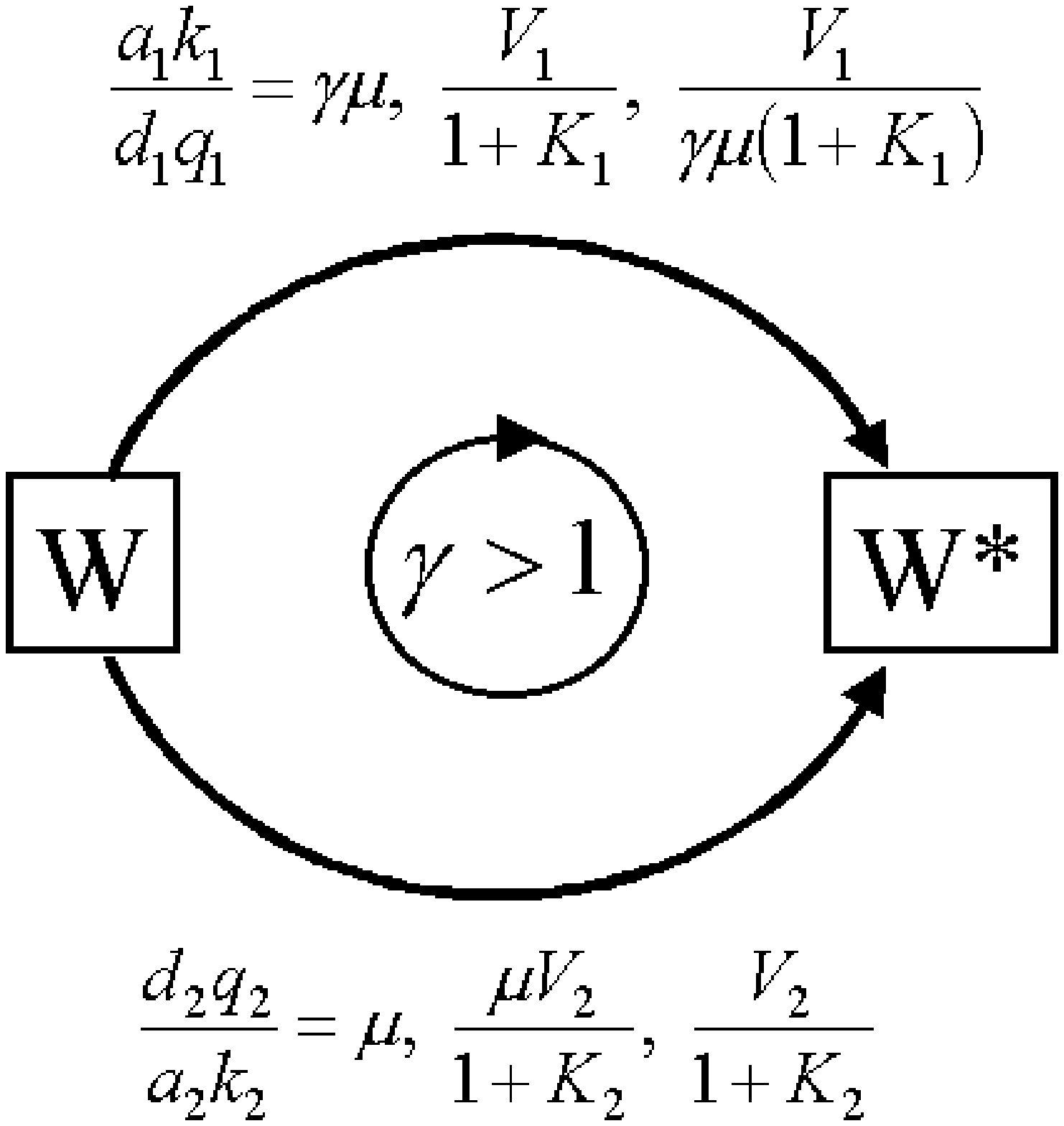,%
width=3.in,height=4.in,%
bbllx=1.5in,bblly=2.in,%
bburx=7.in,bbury=9in}
\]
\caption{A semi-quantitative, Michaelis-Menten, representation for 
the PdPC. The three numbers by each pathway are the equilibrium 
constant, forward and backward rates for the enzymatic reaction.  
They satisfy their respective Haldane relationship for 
thermodynamic reversibility.  In general the steady-state 
$\frac{[W^*]}{[W]}$ is between $\mu$ and $\gamma\mu$.  If 
$V_1\gg V_2$, then it is near $\gamma\mu$, and if $V_2 \gg V_1$,
it is near $\mu$.  When $\gamma > 1$ the the PdPC runs clockwise.}
\end{figure}
When $K_1=K_2$ and $\sigma=\frac{V_1}{V_2}$ $\gg 1$, 
the phosphorylation pathway is dominant.  Hence 
$\frac{[W^*]}{[W]}$ = $\mu\gamma$.  When 
$\sigma\ll 1$, the pathway is dominated by dephosphorylation
and  $\frac{[W^*]}{[W]}$ = $\mu$.
Therefore, for a finite $\gamma$, one does not expect 
$W^*\rightarrow 1$ as $\sigma\rightarrow\infty$, as
clearly pointed out  earlier in \cite{Gresser}.  Rather 
we have $W^*\rightarrow\frac{\mu\gamma}{1+\mu\gamma}$ as 
$\sigma\rightarrow\infty$.  For $\mu=10^3$ and $\gamma$ = 
$10^2$, $10^3$, $10^4$, and $10^{10}$, the plateau of
$W^*$ toward right in Fig. 1 is expected to be 0.099,
$\frac{1}{2}$, $\frac{10}{11}$, and almost $1$. 

	The response coefficient, $R_v$, which 
characterizes the steepness of the transition in covalent 
modification, has been defined as the ratio of the 
$\sigma$ when $W^*$ = 90\% to the $\sigma$ when $W^*$ = 10\%
\cite{GK}.  For a simple Michaelis-Menten kinetics its 
value is 81.  A value of 1 means the transition is 
infinitely steep.  With the finite $\gamma$ and $\mu$, 
in theory, because $W^*$ never exceeds 0.9 for a range 
of $\mu$ and $\gamma$ (Fig. 1), $R_v$ needs to be 
redefined as the ratio of $\sigma$ when $W^*$ = 
$0.9W^*(\infty)+0.1W^*(-\infty)$ to the $\sigma$ when 
$W^*=0.9W^*(-\infty)+0.1W^*(\infty)$, where
$W^*(\infty)$ = $\frac{\mu\gamma}{1+\mu\gamma}$ and 
$W^*(-\infty)$ = $\frac{\mu}{1+\mu}$.  
In physiological reality, $W^*(\infty)>0.9$ and 
$W^*(-\infty)<0.1$; that is $\mu<1/9$ and $\mu\gamma>9$.
Fig. 3 shows how the response coefficient,
\begin{equation}
        R_v = \frac{(\mu-9)(9\mu\gamma-1)(K_1+0.1)(K_2+0.1)}
		{(\mu\gamma-9)(9\mu-1)(K_1+0.9)(K_2+0.9)}
\end{equation}
depends on the phosphorylation potential 
$\Delta G = RT\ln\gamma$.  It is seen that for the 
physiological range of $\Delta G$, the steepness $R_v$ 
reaches its minimal, platuea value given in \cite{GK}. 
\begin{figure}[b]
\[
\psfig{figure=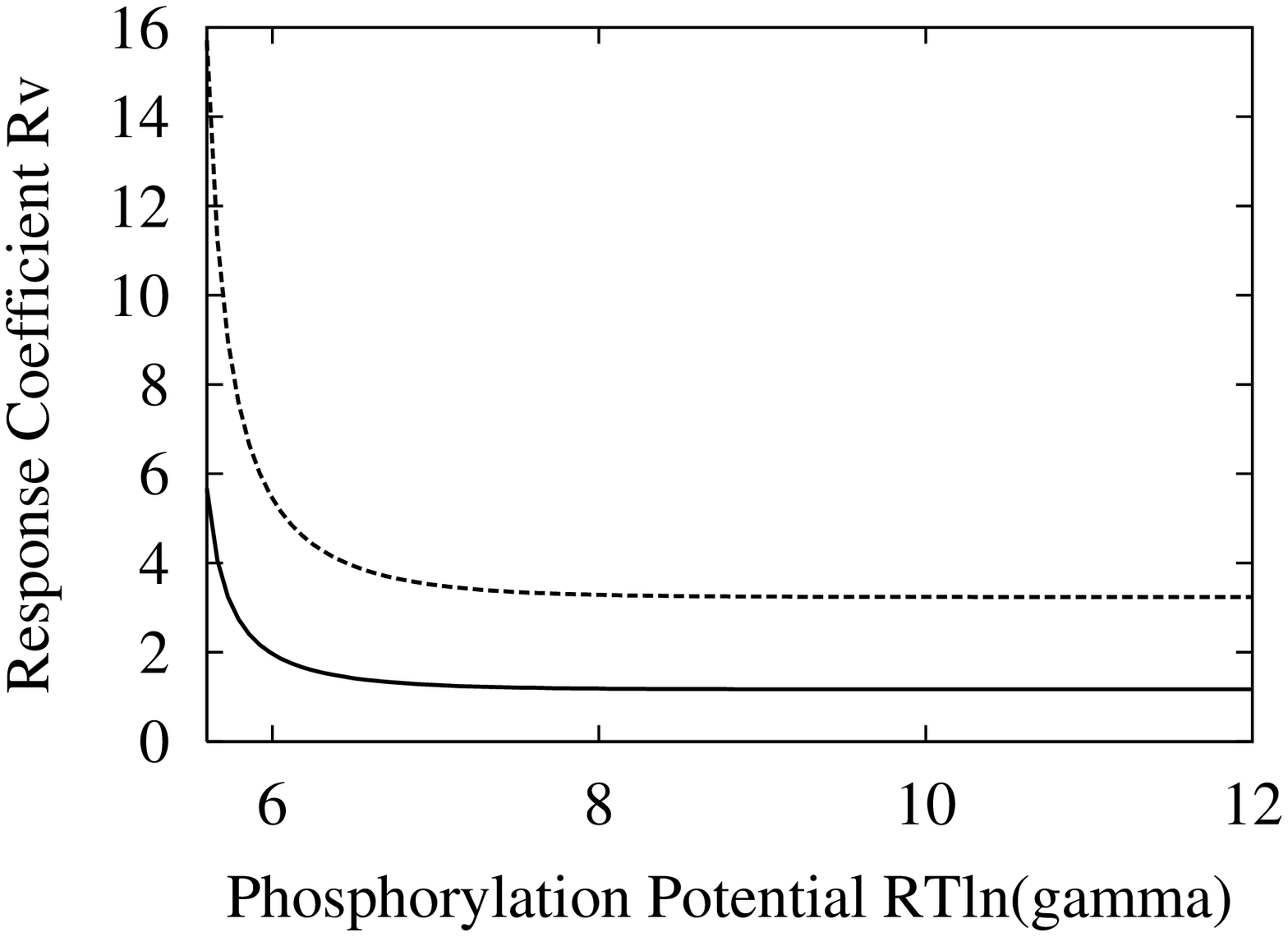,%
width=3.in,height=3.0in,%
bbllx=2.in,bblly=1.in,%
bburx=8.in,bbury=8in,%
angle=-90}
\]
\caption{Intracellular phosphorylation potential, 
$\Delta G = RT\ln\gamma$, in $kcal/mol$, controls
the sensitivity amplification of a PdPC.  The
response coefficient $R_v$ is defined as 
$\sigma(W^*=0.9)/\sigma(W^*=0.1)$ in Fig. 1 \cite{GK}.  
The solid line is for $K_1=K_2=0.01$, and the dashed 
line for $K_1=K_2 =0.1$.  Both with
$\mu=10^{-3}$.}
\end{figure}

	The current model in fact makes a prediction. 
Let $W^*(-\infty)$ and $W^*(\infty)$ be the left and 
right plateaus of the amplification curve in Fig. 1,
which are very close to 0 and 1, respectively.  Then 
\begin{equation}
                 \frac{W^*(-\infty)}{1-W^*(\infty)}
	\approx \frac{W^*(-\infty)}{1-W^*(-\infty)}
	 \times \frac{W^*(\infty)}{1-W^*(\infty)}
	 = \frac{1}{\mu}\times\ \mu\gamma =\gamma. 
\label{nu}
\end{equation}
In contrast, the previous model \cite{GK} predicts
an indeterminate $\frac{0}{0}$.

	The steepness of the curves in Fig. 1 can also be
characterized by the slope at its mid-point, known as 
Hill coefficient in the theory of allosteric cooperativity
\cite{Str}.  It can be obtained analytically from 
Eq. \ref{sigma}:
\begin{equation}
      n_v=\left(\frac{dW^*}{d\ln\sigma}\right)_{W^*=0.5}
	 \approx \frac{1}{4}
	\left(\mu+\frac{1}{\mu\gamma}+K_1+K_2\right)^{-1}
\label{nv}
\end{equation}
when $K_1$, $K_2$, and $\mu$ are small and $\mu\gamma$ 
is large.  We see again that the steepness increases
with increasing $\gamma$.

\section{Temporal Cooperativity}

	Allosteric change in and covalent modification of
proteins are two most basic phenomena in cellular
signaling processes \cite{KGS}.  While the equilibrium 
thermodynamic principle of the former is well understood 
\cite{WG}, relatively little attention has been given to the 
nonequilibrium steady-state thermodynamics \cite{Hill} of 
the latter.  The analysis developed in the present paper 
indicates that the cooperativity in the cyclic reaction is
temporal, with energy ``stored'' in time rather
than in space as for the allosteric cooperativity.
This concept is similar to the {\it energy relay} which
was first proposed by J.J. Hopfield for understanding the
molecular mechanism of kinetic proofreading in 
protein synthesis \cite{Hop74,Hop80}.  We now elaborate 
on this concept by carrying out a quantitative
comparison between the steady-state system given in
Eq. \ref{rxn} and the allosteric
cooperativity. 

\vskip 0.3cm\noindent
{\bf High-order versus zero-order reactions}

One of the most fundamental difference between allosteric 
cooperativity and zero-order ultrasensitivity is apparently 
the order of the reactions.  Allosteric cooperativity
is based on a reaction with high-order
\begin{equation}
     P+nL \overset{K^n}{\rightleftharpoons} PL_n
\label{allo}
\end{equation}
where $K$ is the equilibrium constant for protein $P$ binding
single-ligand $L$.  The corresponding fraction of protein with 
ligand then is 
\begin{equation}
     Y = \frac{[PL_n]}{[P]+[PL_n]} = \frac{(KL)^n}{1+(KL)^n}.
\label{ac}
\end{equation}
Eq. \ref{ac} indicates that the steepness of the curve
$Y$ versus $\ln(K[L])$ increases with $n$.  On the other
hand, ultrasensitivity is based on both phosphorylation
and dephosphorylation reactions being enzyme limited;
hence both have a very weak dependence on the respective 
substrate concentrations $[W]$ and $[W^*]$.  In the
steady-state
\begin{equation}
          k_{ph}[W]^{\nu} = k_{dp}[W^*]^{\nu}
\label{loworder}
\end{equation}
where $k_{ph}$ and $k_{dp}$ are the rates of phosphorylation and 
dephosphorylation and $\nu$, the ``order of the reaction'', 
is near zero.  (Normally the power term to
the concentration of a species implies the stoichiometry
of that species in a reaction.  The meaning of $\nu$ here
is that the reaction is even less than first-order.
Both a hyperbolic curve, as expected from an enzymatic 
reaction with saturation, and a curve with power $\nu<1$ 
are concave down with negative curvature.)  
The corresponding fraction of protein in the activated 
state
\begin{equation}
    Z = \frac{[W^*]}{[W]+[W^*]} = \frac{k_{ph}^{1/\nu}}
		{k_{ph}^{1/\nu}+k_{dp}^{1/\nu}}.
\label{us}
\end{equation}
Eq. \ref{us} indicates that the steepness of the curve
$Z$ versus $\ln(k_{ph}/k_{dp})$ increases with $1/\nu$.  
Therefore, the optimal situation is a zero-order reaction 
with $\nu=0$. 

	Surprisingly, allosteric binding (Eq. \ref{allo})
can yield an equation identical to (\ref{loworder}).  
Let equilibrium constant $K=k_+/k_-$ where $k_+$ and $k_-$
are association and dissociation rate constants. Then 
in the equilibrium $(k_+[L])^n[P]$ = $k_-^n[PL_n]$.  That
is
\begin{equation}
		k_+[L][P]^{1/n} = k_-[PL_n]^{1/n}.
\label{highorder}
\end{equation}

\vskip 0.3cm\noindent
{\bf Temporal cooperativity in zero-order
reaction cycle}

	The cooperativity achieved by ultrasensitivity, 
therefore, can be stated as follows.  It takes, on average,
$n_v=1/\nu$ PdPCs in order to transform one $W$ to $W^*$. 
There is a temporal cooperativity on the scale of 
$n_v$ cycles.  Therefore, $n_v$ in time is analogous 
to the number of subunits in allosteric cooperativity
(see Eqs. \ref{loworder} and \ref{highorder}).
Most importantly, transforming one $W$ to one $W^*$ through 
multiple ``futile'' cycles is precisely the mechanism proposed
by Hopfield for kinetics proofreading of protein
biosynthesis (with branched reaction pathways) in which 
$n_v\approx 2$ \cite{Hop74,Hop80}.  Multiple branched
pathways have been proposed for kinetic proofreading in
T-cell receptor signaling \cite{McKeithan}.
Of course, the ATP hydrolysis is not futile, rather the 
energy supplies the need to maintain high accuracy and 
sensitivity or improved memory of a steady-state 
``living'' system away from true thermodynamic 
equilibrium.

	The above statement can be further quantified.  Let's 
consider a system with only a single $E_1$ and a single 
$E_2$ molecule, but $n$ $W$ substrate molecules.  The complete
kinetics of $W^*$ formation can be represented by a chain 
kinetic scheme shown in Fig. 4 \cite{QE}, which is a 
detailed version of what is shown in (\ref{rxn}).  Each 
time when a cycle is completed, one ATP molecule is hydrolyzed.
\begin{figure}[b]
\vskip 1.5cm
\[
\psfig{figure=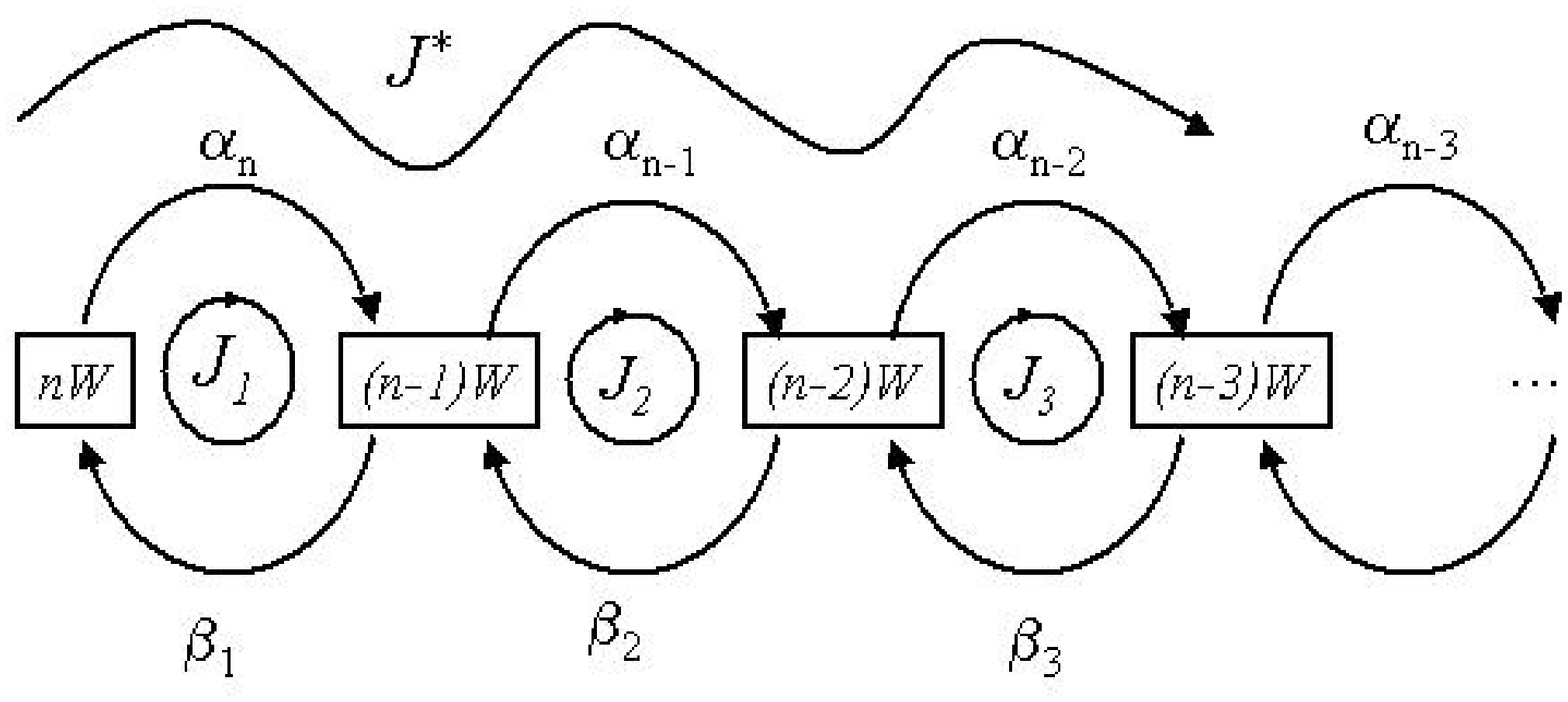,%
width=5.in,height=3.2in,%
bbllx=1.75in,bblly=1.in,%
bburx=7.5in,bbury=10in,%
angle=-90}
\]
\caption{Detailed kinetic scheme showing how the PdPCs are being 
completed while $n$ $W$ molecules are transformed to $W^*$.  
The ``futile'' cycles are indicated by $J_1$, $J_2$, etc., 
and the net flux for $W^*$ formation is denoted by $J^*$. 
According to Michaelis-Menten kinetics, transition rates
$\alpha_i$ = $\frac{k_1E_{1T}}{1+nK_1/i}$ and $\beta_j$ =
$\frac{k_2E_{2T}}{1+nK_2/j}$, which are weakly substrate
dependent when $K$'s are small.  $\frac{\alpha_i}{\beta_j}$
$\neq$ $\frac{i\alpha_1}{j\beta_1}$ means cooperativity.
The process is closely related to a biased random walk 
with $J^*$ and $J_k$ analogous to the ``drift velocity'' 
and ``diffusion constant'', respectively. 
}
\end{figure}
The cooperativity of the kinetics in Fig. 4 is characterized
by 
\begin{equation}
  \frac{\alpha_{n-i}}{\beta_{i+1}}
       \left[\frac{(n-i)\alpha_1}{(i+1)\beta_1}\right]^{-1}
     = \frac{i+1+nK_2}{n-i+nK_1}.
\label{cindex}
\end{equation}
For $n$ completely independent $W$ molecules undergoing 
$W\rightleftharpoons W^*$ transition, Eq. \ref{cindex} 
is expected to be unity.  However the $n$ $W$ molecules 
in Fig 4 are not independent since they are linked by the 
enzymatic reactions.  For small $K_1$ and
$K_2$, there is a cooperative phosphorylation when 
$i>n/2$ and there is a cooperative dephosphorylation when
$i<n/2$.  

	Fig 5 shows the steepness of the response curve 
for the model given in Fig. 4. The detailed model gives the 
same $n_v=12.5$ for $K_1=K_2=0.01$.   The significance 
of this chain model, however, is that it reveals the origin 
of the cooperativity \cite{Hill2}.  Furthermore, according to
the theory of linear cooperativity \cite{PS,Hill2}, the 
steepness of the curves in Fig. 1 is directly related to 
the microscopic fluctuation in the the number of $W^*$. 
\begin{figure}[b]
\vskip 1.5cm
\[
\psfig{figure=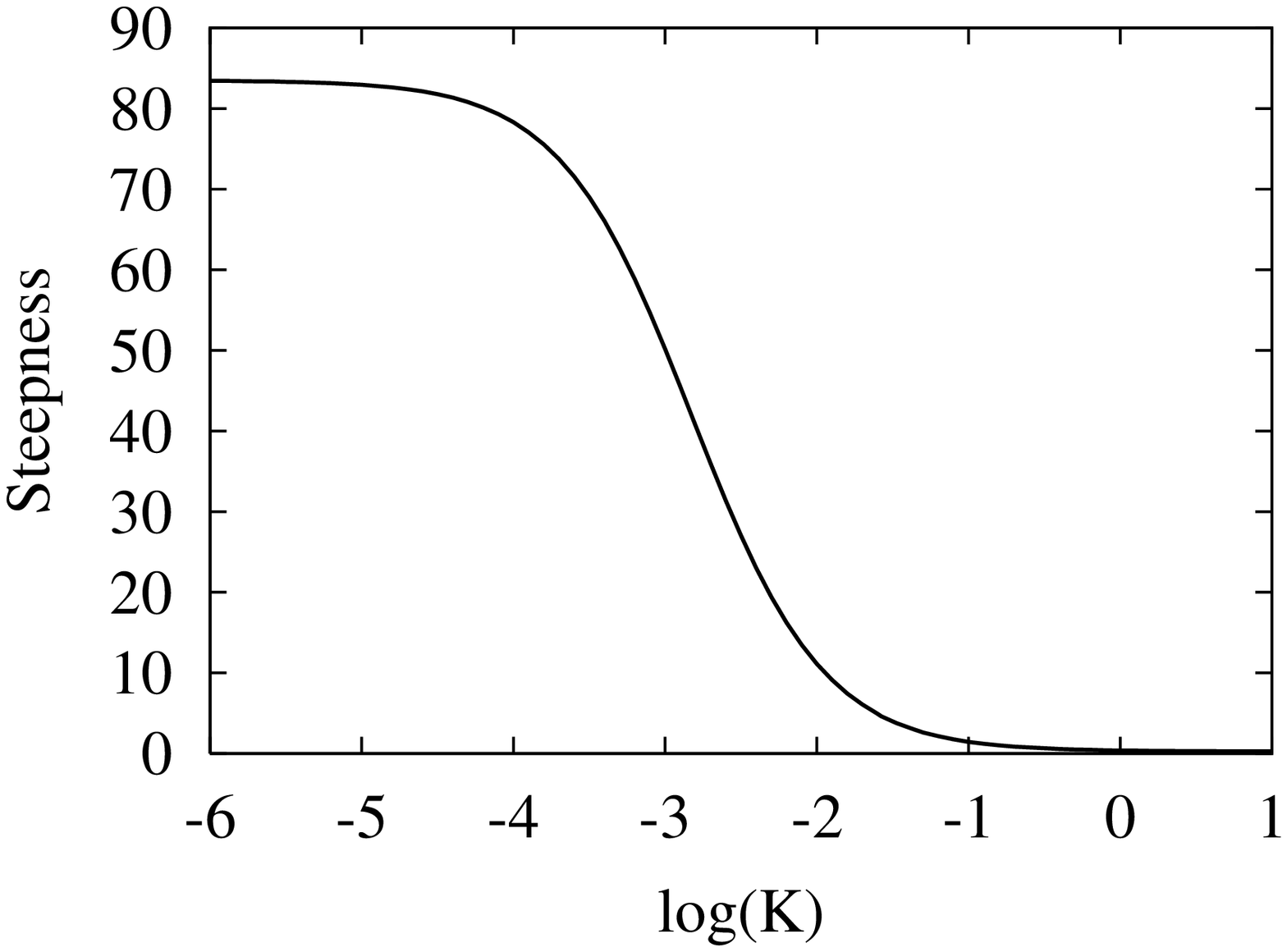,%
width=4.5in,height=3.2in,%
bbllx=2.in,bblly=1.in,%
bburx=7.5in,bbury=9in,%
angle=-90}
\]
\caption{The steepness $n_v$ according to the kinetic model
in Fig 4. First $\frac{[W^*]}{[W]+[W^*]}$ is calculated as
function of $\ln(k_1/k_2)$ with various $K_1=K_2=K$. 
The steepness, $n_v$, of the curve is the slope at 
its mid-point.  Other parameters used: $E_{1T}=E_{2T}=1$, 
$n=1000$.  It can be analytically shown that 
for small $K$, $n_v=(n+2)/12$ = 83.5, and for large $K$, 
$n_v=1/4$.  For $K=0.01$, $n_v \approx 12.5$ according to 
Eq. \ref{nv}. 
}
\end{figure}

	Fig. 6 shows a numerical example of the reaction 
kinetics of the model given in Fig. 4.  The large fluctuations 
in the number of $W^*$ molecules is directly related to the
$n_v$.  In fact, $\sqrt{\langle(\Delta W^*)^2\rangle}$ = 
$\sqrt{n n_v}$ is expected to be 112.  More cooperative system 
has larger fluctuations.
\begin{figure}[b]
\vskip 1.5cm
\[
\psfig{figure=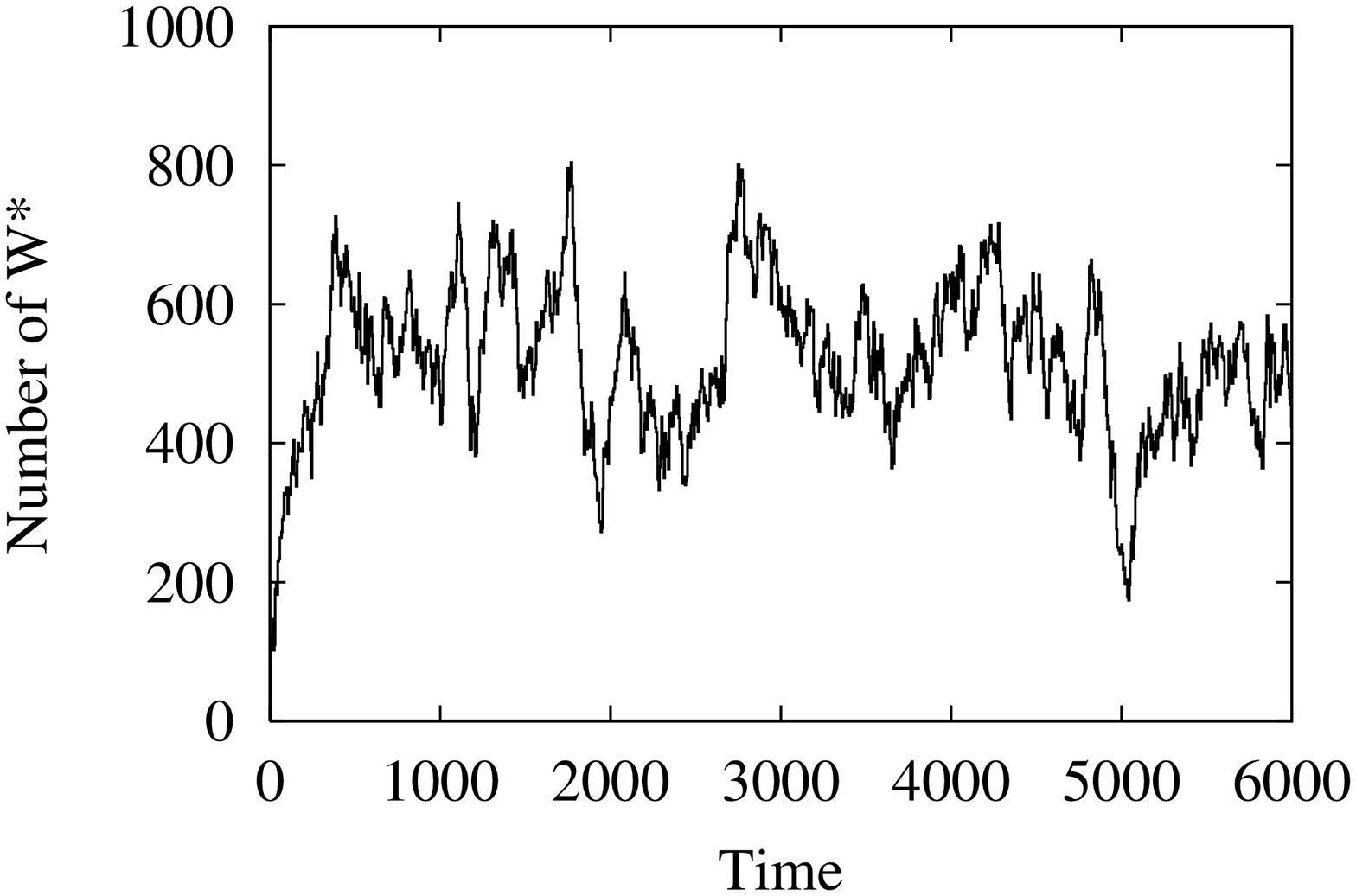,%
width=5.in,height=3.2in,%
bbllx=2.in,bblly=1.in,%
bburx=7.5in,bbury=10in,%
angle=-90}
\]
\vskip -1.cm
\[
\psfig{figure=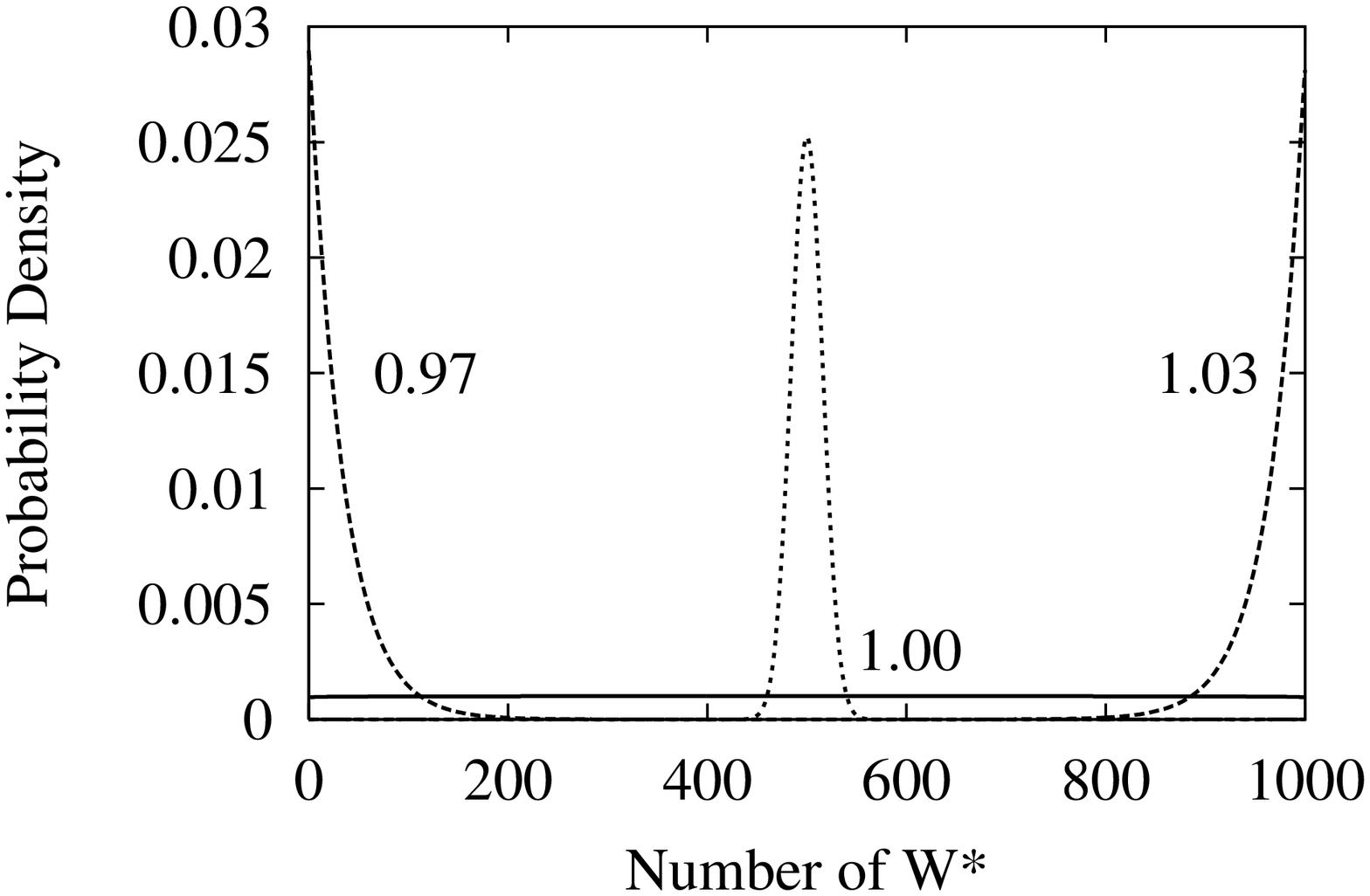,%
width=5.in,height=2.9in,%
bbllx=2.in,bblly=1.in,%
bburx=7.5in,bbury=10.in,%
angle=-90}
\]
\caption{Upper pannel shows a numerical simulation of the reaction 
given in Fig. 4, with $n=1000$, $E_{1T}=E_{2T}=1$, $K_1=K_2=0.01$, 
$k_1=k_2=100$.  Since $\sigma=1$, the steady-state
$[W^*]=500$.  The large fluctuations in the number of
$W^*$ molecules is directly related to the
$n_v$: $\sqrt{\langle(\Delta W^*)^2\rangle}$ = 
$\sqrt{n n_v}$. Lower pannel shows the probability distributions
for the number of $W^*$. Solid flat line: $\sigma=1.00$,
Dashed lines: different distributions for $\sigma=0.97$ and 
$1.03$ respectively.  We see a sharp response to $\sigma$ being 
less and greater than $1$.  In comparison, the central peak with 
dotted line is for non-cooperative system with 1000 independent 
molecules and $\sigma=1$.  More cooperative system has larger 
fluctuations.
}
\end{figure}

\section{Discussion}

	The rigorous thermodynamic analysis of the model
for phosphorylation-dephosphorylation cycle (PdPC)
originally proposed in \cite{SC,GK} indicates that 
a sustained intracellular phosphorylation potential 
is essential in the functioning of the signal 
transduction process.  This result suggests that 
the ubiquitous phosphorylation in biological
signaling processes, in addition to the covalent
chemical modification which leads to structural recognition, 
also utilizes biochemical energy from the high-energy 
phosphate in order to carry out its function with 
high accuracy, robustness, sensitivity, and specificity 
\cite{Hop74,LQ,QBJ}. 
The analysis also reveals a shared mechanism between the 
ultrasensitivity and kinetic proofreading in a large 
class of cellular processes involving GTPases \cite{BSM}.
Both use cycle kinetics \cite{Hill} to improve the power of 
biological selectivity.

	Our quantitative analysis also provided a clear
mechanistic origin for the high cooperativity in the 
zero-order ultrasensitivity.  A chain kinetic model indicates
that the cooperativity is achieved through temporal 
cooperativity.  This mechanism is parallel in mathematical form 
to, but fundamentally different in biochemical nature from, 
the allosteric cooperativity of multi-subunits protein
systems \cite{WG}.  Both temporal and allosteric cooperativities
have a deep connection to the molecular fluctuations as shown
in Fig. 6 \cite{QE}, an insight largely unexplored in the 
studies of biological signal transduction processes.

	In order to compare our result with that of 
Goldbeter and Koshland, we have used the value $K_1=K_2=0.01$ 
in this study.  These values are extreme cases and many
PdPCs studied in laboratory 
experiments show a much less cooperativity.  With 
$K_m \approx$ $0.1-1\mu$M and concentrations of 
$\sim 1\mu$M for the kinases in the MAPK pathway \cite{HF}, 
the realistic value will be $\sim 0.1-1$.   The 
phosphatase concentration is even lower, $\sim 1$nM. 
Note that from Eq. \ref{nv} high cooperativity requires both 
$K$'s for the kinase and the phosphatase to be small. 
The current model analysis also suggests that the source 
of phosphate in a PdPC, while chemically equivalent, could be 
important.  A phosphate from ATP hydrolysis can be energetically 
different from a phosphate from GTP hydrolysis.  In the cells, 
[ATP] $\sim$ 10mM, [ADP] $\sim$ 10$\mu$M, 
[GTP] $\sim$ 1mM,  [GDP] $\sim$ 100$\mu$M, and [Pi] $\sim$ 1mM 
\cite{Kush}.  Therefore, different cellular biochemical 
``batteries'' can have different ``voltages''.

\section{Acknowledgements}
Professor Walter Kauzmann's work on the physical chemistry of 
hydrophobic interactions in proteins was an impotant part of 
my post doctoral education with John Schellman.  In the present work 
I try to follow the same spirit of that work by raising a basic 
physicochemical question about a currently important biological 
problems: Is there a thermodynamic role of the prevalent phosphorylation 
in biological signal transduction processes?  I thank Jon Cooper, 
Eddy Fischer, Bob Franza, John Hopfield, Guangpu Li, and Elliott Ross 
for helpful discussions, and Jim Bassingthwaighte and 
P. Boon Chock for carefully reading the manuscript.

\section{Figure Captions}

\underline{Figure 1.}  
Amplified sensitivity of a phosphorylation-dephosphorylation
cycle as a function of intracellular phosphorylation potential
$\Delta G$.  From top to bottom, $\gamma$ = $10^{10}$, $10^4$,
$10^3$ and $10^2$, corresponding to $\Delta G$= 13.8, 5.5, 4.1, 
and 2.8 $kcal/mol$. $13.8 kcal/mol$ is the typical value for
intracellular phosphorylation potential \cite{Str,Kush}. Other 
parameters used in the computation: $K_1=K_2=0.01$ and $\mu=0.001$.

\vskip 0.3cm
\underline{Figure 2.}  
A semi-quantitative, Michaelis-Menten, representation for 
the PdPC. The three numbers by each pathway are the equilibrium 
constant, forward and backward rates for the enzymatic reaction.  
They satisfy their respective Haldane relationship for 
thermodynamic reversibility.  In general the steady-state 
$\frac{[W^*]}{[W]}$ is between $\mu$ and $\gamma\mu$.  If 
$V_1\gg V_2$, then it is near $\gamma\mu$, and if $V_2 \gg V_1$,
it is near $\mu$.  When $\gamma > 1$ the the PdPC runs clockwise.

\vskip 0.3cm
\underline{Figure 3.}  
Intracellular phosphorylation potential, 
$\Delta G = RT\ln\gamma$, in $kcal/mol$, controls
the sensitivity amplification of a PdPC.  The
response coefficient $R_v$ is defined as 
$\sigma(W^*=0.9)/\sigma(W^*=0.1)$ in Fig. 1 \cite{GK}.  
The solid line is for $K_1=K_2=0.01$, and the dashed 
line for $K_1=K_2 =0.1$.  Both with
$\mu=10^{-3}$.

\vskip 0.3cm
\underline{Figure 4.}  
Detailed kinetic scheme showing how the PdPCs are being 
completed while $n$ $W$ molecules are transformed to $W^*$.  
The ``futile'' cycles are indicated by $J_1$, $J_2$, etc., 
and the net flux for $W^*$ formation is denoted by $J^*$. 
According to Michaelis-Menten kinetics, transition rates
$\alpha_i$ = $\frac{k_1E_{1T}}{1+nK_1/i}$ and $\beta_j$ =
$\frac{k_2E_{2T}}{1+nK_2/j}$, which are weakly substrate
dependent when $K$'s are small.  $\frac{\alpha_i}{\beta_j}$
$\neq$ $\frac{i\alpha_1}{j\beta_1}$ means cooperativity.
The process is closely related to a biased random walk 
with $J^*$ and $J_k$ analogous to the ``drift velocity'' 
and ``diffusion constant'', respectively.

\vskip 0.3cm
\underline{Figure 5.}  
The steepness $n_v$ according to the kinetic model
in Fig 4. First $\frac{[W^*]}{[W]+[W^*]}$ is calculated as
function of $\ln(k_1/k_2)$ with various $K_1=K_2=K$. 
The steepness, $n_v$, of the curve is the slope at 
its mid-point.  Other parameters used: $E_{1T}=E_{2T}=1$, 
$n=1000$.  It can be analytically shown that 
for small $K$, $n_v=(n+2)/12$ = 83.5, and for large $K$, 
$n_v=1/4$.  For $K=0.01$, $n_v \approx 12.5$ according to 
Eq. \ref{nv}.

\vskip 0.3cm
\underline{Figure 6.} 
Upper pannel shows a numerical simulation of the reaction 
given in Fig. 4, with $n=1000$, $E_{1T}=E_{2T}=1$, $K_1=K_2=0.01$, 
$k_1=k_2=100$.  Since $\sigma=1$, the steady-state
$[W^*]=500$.  The large fluctuations in the number of
$W^*$ molecules is directly related to the
$n_v$: $\sqrt{\langle(\Delta W^*)^2\rangle}$ = 
$\sqrt{n n_v}$. Lower pannel shows the probability distributions
for the number of $W^*$. Solid flat line: $\sigma=1.00$,
Dashed lines: different distributions for $\sigma=0.97$ and 
$1.03$ respectively.  We see a sharp response to $\sigma$ being 
less and greater than $1$.  In comparison, the central peak with 
dotted line is for non-cooperative system with 1000 independent 
molecules and $\sigma=1$.  More cooperative system has larger 
fluctuations.


\vskip 0.3cm
\begin{thebibliography}{00} 

\bibitem{K}
D.E. Koshland, The era of pathway quantification, Science, 280 (1998) 852.

\bibitem{HHLM}
L.H. Hartwell, J.J. Hopfield, S. Leibler, A.W. Murray, 
From molecular to modular cell biology, Nature 402 (1999) C47-C52.

\bibitem{Krebs}
E.G. Krebs, Phosphorylation and dephosphorylation of glycogen 
phosphorylase: a prototype for reversible covalent enzyme modification, 
Curr. Top. Cell. Regul. 18 (1981) 401-419. 

\bibitem{SC}
E.R. Stadtman, P.B. Chock, Superiority of interconvertible enzyme 
cascades in metabolic regulation: analysis of monocyclic systems,
Proc. Natl. Acad. Sci. USA 74 (1977) 2761-2765.

\bibitem{GK}
A. Goldbeter, D.E. Koshland, An amplified sensitivity arising from 
covalent modification in biological systems, Proc. Natl. Acad. Sci.
USA. 78 (1981) 6840-6844.

\bibitem{KGS}
D.E. Koshland, A. Goldbeter, J.B. Stock, Amplification and adaptation 
in regulatory and sensory systems, Science 217 (1982) 220-225.

\bibitem{SCS}
E. Shacter, P.B. Chock, E.R. Stadtman, Regulation through 
phosphorylation/dephosphorylation cascade systems,
J. Biol. Chem. 259 (1984) 12252-12259.

\bibitem{HF}
C.F. Huang, J.E. Ferrell, Ultrasensitivity in the mitogen-activated 
protein kinase cascade, Proc. Natl. Acad. Sci. USA 93 (1996) 10078-10083. 

\bibitem{FM}
J.E. Ferrell, E.M. Machleder, The biochemical basis of an all-or-none 
cell fate switch in Xenopus oocytes, Science, 280 (1998) 895-898.

\bibitem{GK2}
A. Goldbeter, D.E. Koshland, Energy expenditure in the control of 
biochemical systems by covalent modification, J. Biol. Chem.
262 (1987) 4460-4471. 

\bibitem{LQ}
G.P. Li, H. Qian, Kinetic timing: a novel mechanism for improving 
the accuracy of GTPase timers in endosome fusion and other biological
processes, Traffic 3 (2002) 249-255.

\bibitem{BSM}
H.R. Bourne, D.A. Sanders, F. McCormick, The GTPase superfamiliy:
a conserved switch for diverse cell functions, Nature 348 (1990) 125-131.

\bibitem{Gresser}
M.J. Gresser, Regulation of enzyme activity by cyclic 
phosphorylation-dephosphorylation cascades. Thermodynamic
constraints, Biochim. Biophys. Acta 743 (1983) 316-322. 

\bibitem{Str}
L. Stryer, Biochemistry, W.H. Freeman, San Francisco, 1981.

\bibitem{Kush}
M.J. Kushmerick, Energy balance in muscle activity: simulations of 
ATPase coupled to oxidative phosphorylation and to creatine kinase,
Compara. Biochem. Physiol. B. 120 (1998) 109-123.

\bibitem{WG}
J. Wyman, S.J. Gill, Binding and Linkage: Functional Chemistry 
of Biological Macromolecules, University Science Books, Herndon, 
VA, 1990.

\bibitem{Hill}
T.L. Hill,  Free Energy Transduction in Biology:
The Steady-State Kinetic and Thermodynamic Formalism,
Academic Press, New York, 1977.

\bibitem{Hop74}
J.J. Hopfield, Kinetic proofreading: a new mechanism for reducing
errors in biosynthetic processes requiring high specificity,
Proc. Natl. Acad. Sci. USA 71 (1974) 4135-4139.

\bibitem{Hop80}
J.J. Hopfield, The energy relay: a proofreading scheme based on 
dynamic cooperativity and lacking all characteristic symptoms of 
kinetic proofreading in DNA replication and protein synthesis,
Proc. Natl. Acad. Sci. USA 77 (1980) 5248-5252.

\bibitem{McKeithan}
T.W. McKeithan, Kinetic proofreading in T-cell receptor signal
transduction, Proc. Natl. Acad. Sci. USA 92 (1995) 5042-5046.

\bibitem{QE}
H. Qian, E.L. Elson, 
Single-molecule enzymology: stochastic Michaelis-Menten kinetics,
Biophys. Chem. 101 (2002) 565-576.

\bibitem{Hill2}
T.L. Hill, Cooperativity Theory in Biochemistry:
Steady-State and Equilibrium Systems, Springer-Verlag, 
New York, 1985. 

\bibitem{PS}
D. Poland, H.A. Scheraga, Theory of Helix-Coil Transitions,
Academic Press, New York, 1970.

\bibitem{QBJ}
H. Qian, Amplifying signal transduction specificity without
multiple phosphorylation, Biophys. J. in the press.


\end{thebibliography}
\end{document}